\documentclass[aps,pre,twocolumn,floats,floatfix,english]{revtex4}
\usepackage[T1]{fontenc}
\usepackage[latin1]{inputenc}
\usepackage{amsmath}
\usepackage{setspace}
\usepackage{amssymb}

\makeatletter

\providecommand{\LyX}{L\kern-.1667em\lower.25em\hbox{Y}\kern-.125emX\@}

\usepackage{setspace}
\usepackage{epsfig}
\usepackage{babel}
\makeatother

\begin{document}
\title{Asset trees and asset graphs in financial markets}

\author{J.-P. Onnela, A. Chakraborti}
\author{K. Kaski}
\email{Kimmo.Kaski@hut.fi}
\affiliation{Laboratory of Computational Engineering, 
Helsinki University of Technology, P.O. Box 9203, FIN-02015 HUT, Finland}
\author{J. Kertész}
\affiliation{Department of Theoretical Physics, Budapest University 
of Technology \& Economics, Budafoki út 8, H-1111, Budapest, Hungary}
\affiliation{Laboratory of Computational Engineering, 
Helsinki University of Technology, P.O. Box 9203, FIN-02015 HUT, Finland} 
\author{A. Kanto}
\affiliation{Department of Quantitative Methods in Economics and 
Management Science, Helsinki School of Economics, P.O.Box 1210, 
FIN-00101 Helsinki, Finland}

\begin{abstract}

This paper introduces a new methodology for constructing a network of companies called a dynamic asset graph. This is similar 
to the dynamic asset tree studied recently, as both are based on correlations between asset returns.
However, the new modified methodology does not, in general, lead to a tree but a graph, or several 
graphs that need not be inter-connected. 
The asset tree, due to the minimum spanning tree criterion, is forced to ``accept'' edge lengths that 
are far less optimal (longer) than the asset graph, thus resulting in higher overall length for the tree. 
The same criterion also causes asset trees to be more fragile in structure when measured by the single-step survival ratio.
Over longer time periods, in the beginning the asset graph decays more slowly than the asset tree, 
but in the long-run the situation is reversed. 
The vertex degree distributions indicate that the possible scale free behavior of the asset graph is not as evident as it is 
in the case of the asset tree.

\end{abstract}

\maketitle 

\section{Introduction}

Mantegna suggested studying the clustering of companies using the
correlation matrix of asset returns \cite{Man1}. The correlations are
transformed into distances, a subset of which is selected using the
minimum spanning tree (MST) criterion. In the resulting tree, the
distances are the edges used to connect the nodes, or companies, and
thus a taxonomy of the financial market is formed.  This method was later
studied by Bonanno et al. \cite{Gio2}, while other studies on clustering
in the financial market are \cite{Kull,Gio,Kas,Lal,Mar}, and those
specifically on market crashes \cite{Lillo,Lillo2}.

Recently, we have studied the properties of a set of asset trees created using the methodology introduced by 
Mantegna in \cite{jpo,short,long}, and applied it in the crash context in \cite{bali}.
In these studies, the obtained multitude of trees was interpreted as a sequence of evolutionary steps of a 
single `dynamic asset tree', and different measures 
were used to characterize the system, which were found to reflect the state of the market. 
The economic meaningfulness of the emerging clustering was also discussed and the dynamic asset trees were found 
to have a strong connection to portfolio optimization. 

In this paper, we introduce a modified methodology which, in general, does not not lead to a tree but a graph, 
or possibly even several 
graphs that do not need to be inter-connected. Here we limit ourselves to studying only one type 
of 'dynamic asset graph', which in terms of its size 
is compatible and thus comparable with the dynamic asset tree. 
Although in graph theory a tree is defined as a type of graph, 
the terms asset graph and asset tree are used here to refer to the two different approaches and 
as concepts are mutually exclusive and noninterchangeable.
The aims of this paper are to introduce this 
modified approach and to demonstrate some of its similarities and 
differences with the previous approach. 
Further considerations of topology and taxonomy of the financial market obtained using dynamic asset graphs 
are to be presented later.

\section{Constructing graphs and trees}

In this paper, the term financial market refers to 
a set of asset price data commercially available from the Center for Research 
in Security Prices (CRSP) of the University of Chicago Graduate School 
of Business. Here we will study the split-adjusted daily closure prices 
for a total of $N=477$ stocks traded at the New York Stock Exchange (NYSE) 
over the period of 20 years, from 02-Jan-1980 to 31-Dec-1999. This 
amounts a total of 5056 price quotes per stock, indexed by time variable 
$\tau = 1, 2, \ldots, 5056$. For analysis and smoothing purposes, the 
data is divided time-wise into $M$ \emph{windows} $t=1,\, 2,...,\, M$ 
of width $T$, where $T$ corresponds to the number of daily returns included in 
the window. Several consecutive windows overlap with each other, the 
extent of which is dictated by the window step length parameter 
$\delta T$, which describes the displacement of the window and is also measured  
in trading days. The choice of window width is a trade-off between too 
noisy and too smoothed data for small and large window widths, 
respectively. The results presented in this paper were calculated from 
monthly stepped four-year windows, i.e. $\delta T = 250/12 \approx 21$ days and 
$T=1000$ days. We have explored a large scale of different values for both 
parameters, and the cited values were found optimal \cite{jpo}. With 
these choices, the overall number of windows is $M=195$.

In order to investigate correlations between stocks we first denote 
the closure price of stock $i$ at time $\tau$ by $P_{i}(\tau)$ 
(Note that $\tau$ refers to a date, not a time window). We focus 
our attention to the logarithmic return of stock $i$, given by 
$r_{i}(\tau)=\ln P_{i}(\tau)-\ln P_{i}(\tau-1)$ which for a sequence 
of consecutive trading days, i.e. those encompassing the given window 
$t$, form the return vector $\boldsymbol r_{i}^t$. In order to 
characterize the synchronous time evolution of assets, we use the equal 
time correlation coefficients between assets $i$ and $j$ defined as

\begin{equation}
\rho _{ij}^t=\frac{\langle \boldsymbol r_{i}^t \boldsymbol r_{j}^t \rangle -\langle \boldsymbol r_{i}^t \rangle \langle \boldsymbol r_{j}^t \rangle }{\sqrt{[\langle {\boldsymbol r_{i}^t}^{2} \rangle -\langle \boldsymbol r_{i}^t\rangle ^{2}][\langle {\boldsymbol r_{j}^t}^{2} \rangle -\langle \boldsymbol r_{j}^t \rangle ^{2}]}},
\end{equation}

\noindent where $\left\langle ...\right\rangle $ indicates a time 
average over the consecutive trading days included in the return 
vectors. Due to Cauchy-Schwarz inequality, these correlation 
coefficients fulfill the condition $-1\leq \rho _{ij}\leq 1$ 
and form an $N\times N$ correlation matrix $\mathbf{C}^t$, which 
serves as the basis for graphs and trees discussed in this paper.


For the purpose of constructing asset graphs and asset trees, 
we define a distance between a pair of stocks.
This distance is associated with the edge connecting the 
stocks and it is expected to reflect the level at which the stocks are correlated.
We use a simple non-linear transformation $d^t_{ij}=\sqrt{2(1-\rho _{ij}^t)}$ to 
obtain distances with the property $2\geq d_{ij}\geq 0$, forming an 
$N\times N$ symmetric distance matrix $\mathbf{D}^t$. 
Now two alternative approaches may be adopted. The first one leads to asset trees, 
and the second one to asset graphs. In both approaches the trees (or graphs) for different time windows are not 
independent of each other, but form a series through time. 
Consequently, this multitude of trees or graphs is interpreted as a 
sequence of evolutionary steps of a single \emph{dynamic asset tree} or 
\emph{dynamic asset graph}.

In the first approach we construct an asset tree according to the methodology by Mantegna \cite{Man1}. 
This approach requires an additional hypothesis about the topology of the metric space,
namely, the so-called ultrametricity hypothesis. 
In practice, it leads to determining the minimum spanning tree (MST) of the distances, denoted $\mathbf{T}^t$. 
The spanning tree is a simply connected acyclic (no cycles) graph that connects all $N$ nodes (stocks) 
with $N-1$ edges such that the sum of all edge weights, $\sum _{d_{ij}^t \in \mathbf{T}^t}d_{ij}^t$, is minimum. 
We refer to the minimum spanning tree at time $t$ by the notation $\mathbf{T}^t=(V,E^t)$, where $V$ 
is a set of vertices and $E^t$ is a corresponding set of unordered pairs of vertices, or edges.
Since the spanning tree criterion requires all $N$ nodes to be always present, the 
set of vertices $V$ is time independent, which is why the time superscript has been dropped from notation. 
The set of edges $E^t$, however, does depend on time, as it is expected that edge lengths in the matrix 
$\mathbf{D}^t$ evolve over time, and thus different edges get selected in the tree at different times.

In the second approach we construct asset graphs. This consists of
extracting the $N(N-1)/2$ distinct distance elements from the upper (or
lower) triangular part of the distance matrix $\mathbf{D}^t$, and
obtaining a
sequence of edges $d_{1}^t, d_{2}^t, \ldots, d_{N(N-1)/2}^t$, where we
have used a single index notation. The edges are then sorted in an
ascending order and form an ordered sequence $d_{(1)}^t, d_{(2)}^t,
\ldots, d_{(N(N-1)/2)}^t$.  Since we require the graph to be
representative of the market, it is natural to build the graph by
including only the strongest connections in it. The number of edges to
include is, of course, arbitrary.  Here we include only $N-1$ shortest
edges in the graph, thus giving $V^t=\{d_{(1)}^t, d_{(2)}^t, \ldots,
d_{(N-1)}^t\}$.  This is motivated by the fact that the asset tree also
consists of $N-1$ edges, and this choice renders the two methodologies
comparable, and possibly even similar to one another. The presented
mechanism
for constructing graphs defines them uniquely and, consequently, no
additional hypotheses about graph topology are required.  It is important
to note that both the set of vertices $V^t$ and the set of edges $E^t$ are
time dependent, and thus we denote the graph with
$\mathbf{G}^t=(V^t,E^t)$. The choice to include only the $N-1$ shortest
edges in the graph means that the \emph{size} of the graph, defined as the
number of its edges, is fixed at $N-1$.  However, the \emph{order} of the
graph, defined as the number of its vertices, is not fixed for the graph
but varies as a function of time. This is due to the fact that even a
small set of vertices may be strongly inter-connected, and thus may use up
many of the available edges. This may also lead to the formation of loops
in the graph. These aspects are clearly different from the tree approach,
where the order is always fixed at $N$ and no loops are allowed by
definition.

\section{Market characterization}

\begin{figure}
\epsfig{file=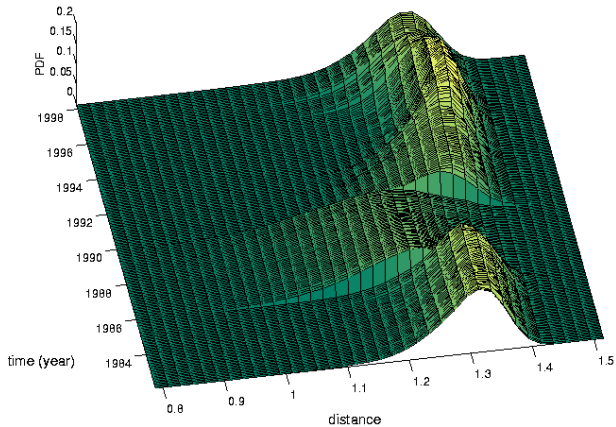,width=3.2in }
\caption{Distribution of all $N(N-1)/2$ distance elements $d_{ij}$ contained in the distance matrix $\mathbf{D}^t$ as a function of time.}
\label{all_distances}
\end{figure}

\begin{figure}
\epsfig{file=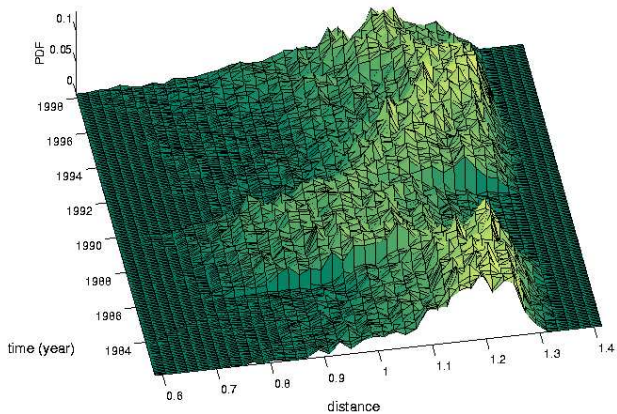,width=3.2in }
\caption{Distribution of the $(N-1)$ distance elements $d_{ij}$ contained in the asset (minimum spanning) tree $\mathbf{T}^t$ as a function of time.}
\label{mst_distances}
\end{figure}

\begin{figure}
\epsfig{file=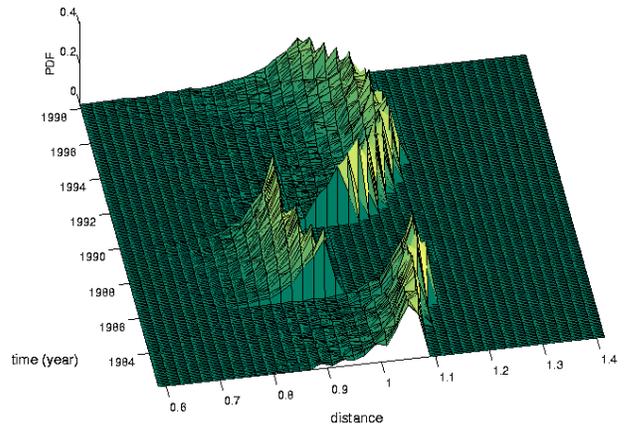,width=3.2in }
\caption{Distribution of the $(N-1)$ distance elements $d_{ij}$ contained in the asset graph $\mathbf{G}^t$ as a function of time.}
\label{graph_distances}
\end{figure}

We start comparisons between the two approaches by visually examining edge length distribution plots. 
The three plots are the distribution of (i) distance elements $d^t_{ij}$ contained in 
the distance matrix $\mathbf{D}^t$ (Figure \ref{all_distances}), (ii) distance elements $d_{ij}$ 
contained in the asset (minimum spanning) tree $\mathbf{T}^t$ (Figure \ref{mst_distances}), and 
(iii) distance elements $d_{ij}$ contained in the asset graph $\mathbf{G}^t$ (Figure \ref{graph_distances}). 
In all three plots, but most prominently in Figure \ref{all_distances}, 
there appears to be a discontinuity in the distribution between roughly 1986 and 1990. It seems that part 
has been cut out, pushed to the left and made flatter. This anomaly is a manifestation of 
Black Monday (October 19, 1987), and its length along the time axis is related to the choice of window width $T$ \cite{bali,long}. 

We can now reconsider the tree and graph construction mechanism described earlier.
Starting from the distribution of the $N(N-1)/2$ distance matrix elements in Figure \ref{all_distances}, 
for the asset tree we pick the shortest $N-1$ of them, subject to the constraint that all vertices are spanned by the 
chosen edges. For the purpose of building the graph, however, this constraint is dropped and we pick the shortest elements 
from the distribution in Figure \ref{all_distances}. Therefore, the distribution of graph edges in Figure \ref{graph_distances} is 
simply the left tail of the distribution of distance elements in Figure \ref{all_distances}, and it seems that 
the asset graph rarely contains edges longer than about 1.1, and the largest distance element is $d_{max}=1.1375$. 
In contrast, in the distribution of tree edges in Figure \ref{mst_distances} most edges included in the tree seem to come 
from the area to the right of the value 1.1, and the largest distance element is now $d_{max}=1.3549$.

\begin{figure}
\epsfig{file=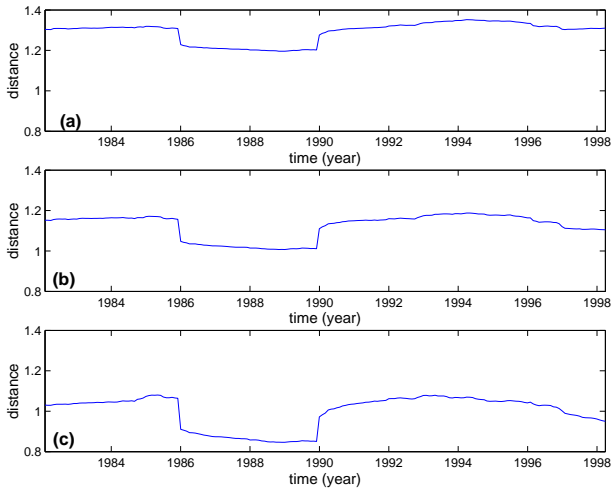,width=3.2in }
\caption{
(a) Mean distance coefficient averaged over all distance elements. 
(b) Normalized tree length average over the edges included in the tree using the MST criterion. 
(c) Normalized graph length averaged over the edges included in the graph by including the shortest edges in the graph.}
\label{av_distances}
\end{figure}

Instead of using the edge length distributions as such, we can characterize the market by studying the location (mean) 
of the edge length distribution by defining a simple measure, the \emph{mean distance}, as 

\begin{equation}
\bar{d}(t)=\frac{1}{N(N-1)/2}\sum _{d_{ij}^t \in \mathbf{D}^t}d_{ij}^t,
\end{equation}

\noindent where $t$ denotes the time at which the tree is 
constructed, and the denominator is the number of distinct elements in the matrix. 
We mention in passing that one could instead use the \emph{mean correlation coefficient}, defined as 

\begin{equation}
\bar{\rho}(t)=\frac{1}{N(N-1)/2}\sum _{\rho_{ij}^t \in \mathbf{C}^t}\rho _{ij}^t, 
\end{equation}

\noindent which would lead to the same conclusions, as the mean distance and mean correlation coefficient are mirror 
images of one another and, consequently, it suffices to examine either one of them. 
In a similar manner, we can characterize the asset tree and the asset graph, which are both simplified networks representing the market, but use only $N-1$ distance elements $d_{ij}^t$ out of the available $N(N-1)/2$ in the distance matrix $\mathbf{D}^t$. 
We define the \emph{normalized tree length} for the asset tree as

\begin{equation}
L_\text{mst}(t)=\frac{1}{N-1}\sum _{d_{ij}^t \in \mathbf{T}^t}d_{ij}^t,
\end{equation}

\noindent and the \emph{normalized graph length} for the asset graph as

\begin{equation}
L_\text{graph}(t)=\frac{1}{N-1}\sum _{d_{ij}^t \in \mathbf{G}^t}d_{ij}^t,
\end{equation}

\noindent where $t$ again denotes the time at which the tree or graph is
constructed, and $N-1$ is the number of edges present.  The three measures
are depicted in Figure \ref{av_distances}.  Three observations are made.
First, all three behave very similarly, which is also reflected by their
level of mutual correlation.  Pearson's linear and Spearman's rank-order
correlation coefficients between the mean distance and normalized tree
length are 0.98 and 0.92, respectively. The same measures between the mean
distance and the normalized graph length are 0.96 and 0.87. Thus, the
normalized tree length seems to track the market slightly better. Second,
the measures decrease in moving from the mean distance, via the normalized
tree length, to the normalized graph length, and their averages are 1.29,
1.12 and 1.00, respectively. Also, the normalized tree length is always
higher than the the normalized graph length, and the difference is 0.13 on
average. It seems that the asset tee, due to the minimum spanning tree
criterion, is forced to ``accept'' edge lengths that are far less optimal
(longer) than the asset graph, resulting in a higher average value for the
normalized tree length than for the normalized graph length.  Third, the
normalized graph length tends to exaggerate the depression caused by the
crash, which can be traced back to the graph construction
mechanism.  Although all of the above measures are very simple, we have
earlier studied the normalized tree length and found it to be descriptive
of the overall market state. It is closely related to market
diversification potential, i.e. the scope of the market to eliminate
specific risk of the minimum risk Markowitz portfolio \cite{jpo,short}.
The fact that the normalized distance and normalized tree length behave so
similarly suggests that they are, at least to some extent, interchangeable
measures.


\begin{figure}
\epsfig{file=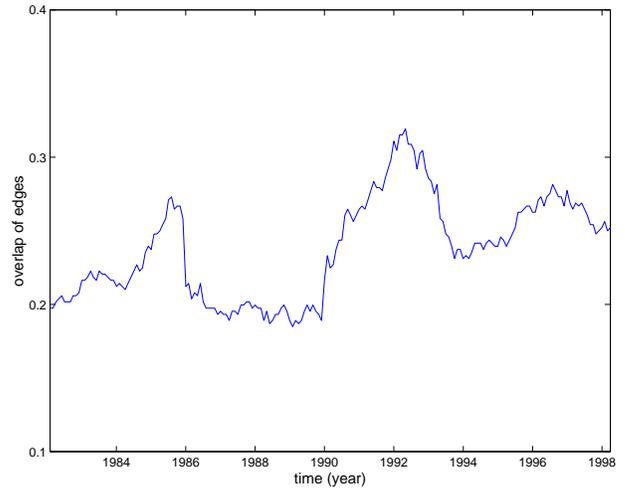,width=3.2in }
\caption{Overlap of edges in the asset graph $\mathbf{G}^t$ and the asset tree $\mathbf{T}^t$ as a function of time.}
\label{overlap_vs_time}
\end{figure}

Based on how the asset graphs and trees are constructed, the edge length distributions in 
Figures \ref{mst_distances} and \ref{graph_distances}, and the normalized graph and 
tree length plots in Figure \ref{av_distances}, 
it is evident that they consist of different edges and, consequently, have different topologies. 
We can learn about the overall topological differences between the asset graph $\mathbf{G}^t=(V_G^t,E_G^t)$ and  
the asset tree $\mathbf{T}^t=(V_T,E_T^t)$ by studying the overlap of edges present in both as a function of time. 
The relative overlap is given by the quantity $\frac{1}{N-1}|E_G^t \cap   E_T^t|$ where $\cap $ is 
the intersection operator and $|...|$ gives the number of elements in the set. As can be see from the plot 
in Figure \ref{overlap_vs_time}, the quantity is quite stable over time. On average, the asset graph and asset tree 
seem to share some 25\% of edges.

\begin{figure}
\epsfig{file=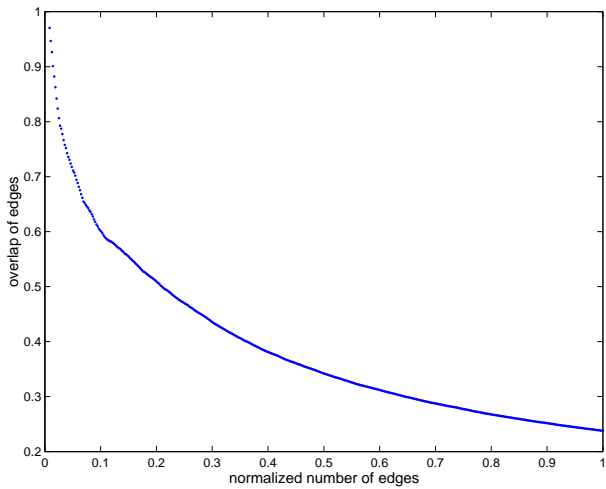,width=3.2in }
\caption{Overlap of edges in the asset graph $\mathbf{G}^t$ and the asset tree $\mathbf{T}^t$ as a function of normalized 
number of edges, average of the time dimension.}
\label{overlap_vs_edges}
\end{figure}

It is also of interest to study how this overlap of edges changes in the process of determining the asset graph and tree. 
Assume we use Kruskal's algorithm to obtain the minimum spanning tree when constructing the asset tree. 
We would first arrange the edges in nondecreasing order by 
their length as a list, and select the shortest unexamined edge for inclusion in the tree, with the condition that it does 
not form a cycle. If it does, we discard it, and move on to the next unexamined edge on the list. Except for the constraint on 
cycles, the algorithm is identical to the way asset graphs are generated. 
If we denote the size of graph \emph{in construction} 
by $n$, where $n=1,2, \ldots, N-1$, then at least for small values of $n$ asset graphs and asset trees should contain 
the same set of edges, i.e., $E(n)_G^t = E(n)_T^t$ and, therefore, be identical in topology.
It is expected that, starting from some value of $n=n_c$, the above equality no longer holds. 
This indicates the formation of the first cycle and, consequently, for all $n \ge n_c$ the asset graph and tree are topologically 
different. 
This is demonstrated in Figure \ref{overlap_vs_edges}, where the relative overlap of edges, $\frac{1}{N-1}|E_G^t \cap   E_T^t|$, 
has been plotted as a function of normalized number of edges, $\frac{n}{N-1}$, and the quantity has been averaged over time. 
The function decreases rapidly for small values of $\frac{n}{N-1}$, indicating that for the current set of data with $N=477$, 
only a few edges can be added before the first cycle is formed.

\section{Evolution of asset graphs and trees}


The robustness of asset graph and asset tree topology can be studied
through the concept of 
\emph{single-step survival ratio}, defined as the fraction of edges found common in two consecutive 
graphs or trees at times $t$ and $t-1$ as

\begin{equation}
\sigma (t)=\frac{1}{N-1}|E^t \cap E^{t-1}|.
\end{equation}

\begin{figure}
\epsfig{file=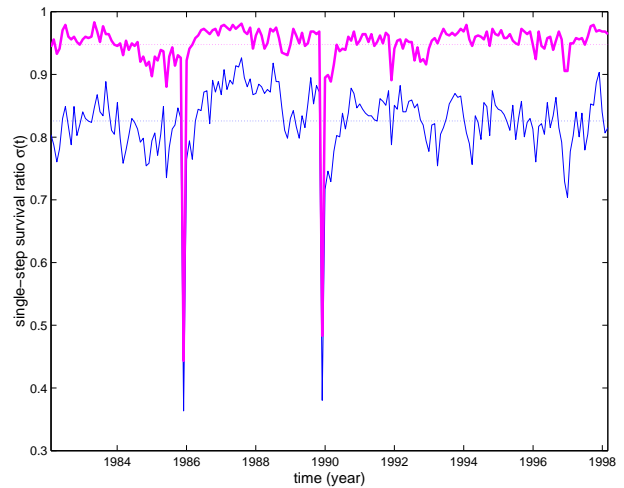,width=3.2in }
\caption{Single-step survival ratios $\sigma (t)$ as functions of time. 
The thicker (upper) curve is for the graph and the thinner (lower) for the tree. 
The dashed lines indicate the corresponding average values.}
\label{sss}
\end{figure}

\noindent As before, $E^t$ refers to the set of edges of the graph or the  
tree 
at time $t$, $\cap $ is the intersection operator and $|...|$ gives 
the number of elements in the set. Although it has not been explicitly indicated 
in the definition, $\sigma(t)$ is dependent on the two parameters, 
namely, the window width $T$ and the step length $\delta T$. 
Figure \ref{sss} shows the plots of single-step survival for both the graph (upper curve) and the tree (lower curve) 
for $\delta T = 250/12 \approx 21$ days and $T=1000$ days. 

The most evident observation is that the graph seems to have a higher
survival ratio than the tree. For the graph, on average, some 94.8\% of
connections survive, whereas the corresponding number for the tree is
about 82.6\%. In addition, the fluctuation of the single-step survival
ratio, as measured by its standard deviation, is smaller for the graph at
5.3\% than for the tree at 6.2\%. In general, both curves fluctuate
together, meaning that the market events causing rewiring in the graph
also cause re-wirings in the tree. This is very clear in the two sudden
dips in both curves, which result from the re-wirings related to Black
Monday
\cite{bali}. Although both curves fall drastically, the one for the asset
graph falls less, indicating that the graph is more stable than the tree
also under extreme circumstances, such as market crashes. The higher
survival ratio for the asset graph is not caused by any particular choice
of parameters, but is reproduced for all examined parametric values. Some
indication of the sensitivity of the single-step survival ratio on the
window width parameter is given in Table \ref{sss_table}.  
The fact that
asset graphs are more stable than asset trees is related to their
construction mechanism. The spanning tree constraint practically never
allows choosing the shortest available edges for the tree and,
consequently, the ensuing structure is more fragile. A short edge between
a pair of stocks corresponds to a very high correlation between their
returns. This may result from the companies having developed a cooperative
relationship, such as a joint venture, or then may be simply incidental.
In the first case the created bond between the two stocks is likely to be
longer lasting than if the MST criterion forced us to include a weaker
bond between two companies.

\begin{table}
\begin{tabular}{|l|l|c|c|c|}
\hline
 &      & T=500 & T=1000 &  T=1500  \\
\hline
mean $\sigma (t)$ & tree  & 72.3\% & 82.6\% & 86.9\% \\
                  & graph & 90.1\% & 94.8\% & 96.0\% \\
std $\sigma (t)$  & tree  & 7.5\% & 6.2\% & 5.4\% \\
                  & graph & 6.3\% & 5.3\% & 4.5\% \\
\hline
\end{tabular}
\caption{Mean and standard deviation of the single-step survival ratio $\sigma (t)$ for the asset graph and the asset tree 
for different values of window width $T$.}
\label{sss_table}
\end{table}

\begin{figure}
\epsfig{file=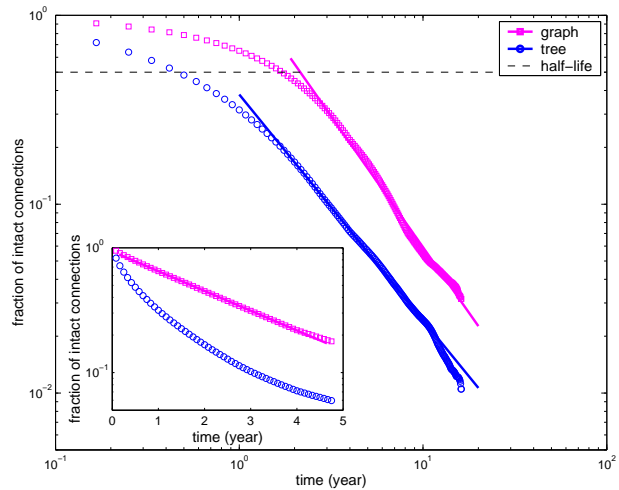,width=3.2in }
\caption{Multi-step survival ratio $\sigma (t,k)$ for asset graph and tree as a function of survival time $k$, 
averaged over the time domain $t$.}
\label{mss}
\end{figure}

We can expand the concept of single-step survival ratio to cover survival over several consecutive time steps $\delta T$. 
Whereas the single-step survival ratio was used to study short-term survival, or robustness, of graphs and trees, 
the \emph{multi-step survival ratio} is used to study their long-term survival. It is defined as


\begin{equation}
\sigma (t,k) = \frac{1}{N-1} |E^t \cap E^{t-1} \cdots E^{t-k+1} \cap E^{t-k}|,
\end{equation}

\noindent where only those connections that have persisted for the whole time period of length $k \delta T$ 
without any interruptions are taken into account. According 
to this formula, when a bond between two vertices breaks even 
once within $k$ steps and then reappears, it is not counted as a 
survived connection. 
A closely related concept is that of graph or tree \emph{half-life} $t_{1/2}$, defined as the time in which half the number of initial 
connections have decayed, i.e., $\sigma (t,t_{1/2})=0.5$. The multi-step survival ratio is plotted in Figure \ref{mss},
where the half-life threshold is indicated by the dashed horizontal line.

The time axis can be divided into two regions based on the nature of the
decay process, and these regions are located somewhat differently for the
graph and the tree.  The precise locations of the regions are, of course,
subject to speculation, but for the purpose carrying out fits and analysis
they need to be fixed. For the asset graph the first and second regions,
discretized according to $\delta T = \frac{1}{12}$ year as mentioned
before, are defined on the intervals $(\frac{1}{2},4)$ and
$(4\frac{1}{12},16\frac{1}{6})$, respectively, both given in years.  
Within the first region, the graph exhibits clean exponential decay, as
witnessed by the fitted straight line on lin-log scale in the inset of
Figure \ref{mss}. Somewhere in between the two regions there is a
cross-over to power-law behavior, which is evident within the second
region, resulting in a straight line on the log-log scale plot of the same
figure.  For the asset tree the regions are defined on
$(\frac{1}{12},1\frac{1}{2})$ and $(1\frac{7}{12},11\frac{1}{4})$. Within
the first region the asset tree decays faster than exponentially, as can
be
verified by comparison with the straight line decay of the graph in the
inset. Similarly to the tree, there is a cross-over to a power-law,
although the slope is faster than for the graph. If we write the power law
decay as $\left\langle \sigma (t,k) \right\rangle_{t} \sim k^{-\gamma}$,
the fits yield for the asset graph $\gamma \approx 1.39$, whereas for the
asset tree we have $\gamma \approx 1.19$.


The finding concerning the slower decay of the asset graph within the first region is fully compatible with the results obtained 
with single-step survival ratio. Since the graph shows higher survival ratio over a single-step, it is to be expected that, at least 
in the early time horizon (within the first region), graphs should decay more slowly than trees. The half-lives for both the graph 
and the tree occur within the first region, and thus it is not surprising that graph half-life is much longer than tree half-life. 
For the graph, we obtained $t_{1/2} \approx 1.71$ years, and for the tree  $t_{1/2} \approx 0.47$. 
Although the half-lives depend on the value of window width $T$, the differences between them persists for different parametric 
values.
When measured in years, for window widths of $T=2$, $T=4$ and $T=6$, the corresponding half-lives for the asset tree are 0.22, 
0.47 and 0.75 years, whereas for the asset graph they are 0.88, 1.71 and 2.51 years, respectively.
Interestingly, the situation seems to be reversed within the second region, where both decay as power-law. Here the higher exponent 
of $\gamma$ for the asset graph indicates that it actually decays faster than the asset tree. This finding could, of course, be 
influenced by our choice of the window width $T$. Explorations with that parameter revealed another interesting phenomenon. Whereas the 
slope for the asset tree seems to be independent of window width, as discussed in \cite{long}, that of asset graphs is not. 
For $T=500$, $T=1000$ and $T=1500$ we obtained for the asset tree the exponents $\gamma=1.15$, $\gamma=1.19$ and $\gamma=1.17$, respectively, which, within the error bars, are to be considered equal \cite{long}. For the asset graph, we obtain the values of 
$\gamma=2.07$, $\gamma=1.39$ and $\gamma=1.55$. Although no clear trend can be detected in these values, a matter that calls for 
further exploration, it is clear that the value of $\gamma$ is higher for the asset graph than for the asset tree. 
Therefore, the asset graph decays more slowly than the asset tree within the first region, 
while within the second region the situation is just the opposite.


\section{Distribution of vertex degrees in asset graphs and asset trees}

\begin{figure}
\epsfig{file=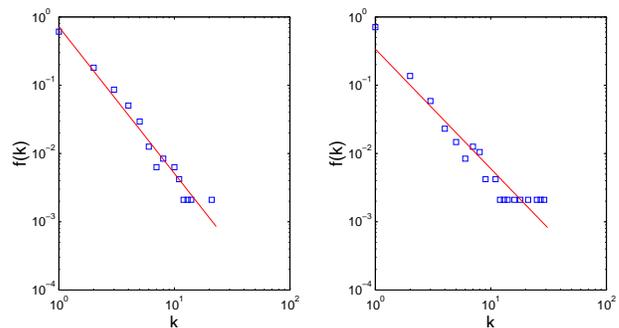,width=3.2in }
\caption{
Typical plots of vertex degree distributions for normal (left) and crash 
topology (right) for the asset tree. The exponents and goodness of fit for them are 
are $\alpha \approx 2.15$, $R^2 \approx 0.96$ and 
$\alpha \approx 1.75$, $R^2 \approx 0.92$.}
\label{mst_vd}
\end{figure}

\begin{figure}
\epsfig{file=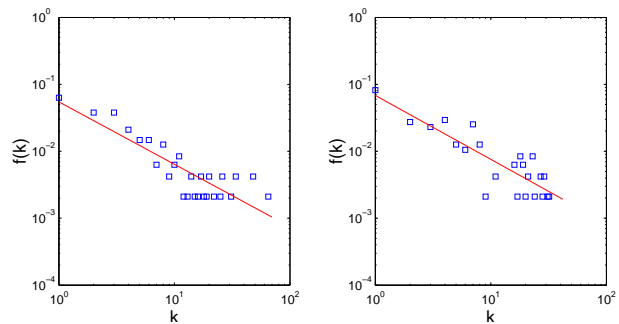,width=3.2in }
\caption{
Typical plots of vertex degree distributions for the asset graph.
The exponents and goodness of fit for them are $\alpha \approx 0.93$, $R^2 \approx 0.74$ and 
$\alpha \approx 0.96$, $R^2 \approx 0.74$, respectively.}
\label{graph_vd}
\end{figure}

As the asset graph and asset tree are representative of the financial market, studying their structure can enhance 
our understanding of the market itself. Recently Vandewalle et al. \cite{Van} found scale free behavior for the asset tree 
in a limited time window. They proposed the distribution of the vertex degrees $f(k)$ to follow a power law of the following form

\begin{equation}
f(k) \sim k ^{-\alpha},
\end{equation}

\noindent with the exponent $\alpha \approx 2.2$. Later, we studied this
phenomenon further with a focus on asset tree dynamics \cite{long}. We
found that the asset tree exhibits, most of the time, scale free
properties with a rather robust exponent $\alpha \approx 2.1 \pm 0.1$
during times of normal stock market operation. In addition, within the
error limits, the exponent was found to be constant over time. However,
during crash periods when the asset tree topology is drastically affected,
the exponent changes to $\alpha \approx 1.8 \pm 0.1$, but nevertheless the
asset tree maintains its scale free character. The interesting question is
whether asset graphs also display similar scale free behavior and if so,
are there are differences in the value of the exponent. As Figures
\ref{mst_vd} and \ref{graph_vd} make clear, for the observed data does not
fit as well with scale free behavior for the asset graph as it does for
the asset tree. The
obtained average value for the exponent of the asset graph is
significantly lower, i.e. $\alpha \approx 0.9 \pm 0.1$. In addition, the
exponent for the aseet graph varies less as a function of time and does
not show distinctively different behavior between normal and crash
markets.

In the case of the asset tree, there were sometimes clear outliers, 
as one node typically had a considerably higher vertex degree than the
power-law scaling would predict. This outlier was used as a \emph{central
node}, a reference against which some tree properties were measured.  
However, the fact that one node often had ``too high'' vertex degree
provided further support for using one of the nodes as the center of the
tree, as discussed in detail in \cite{long}. In the dynamic asset graph,
these types of outliers are not present. This observation merely reflects
upon the differences between the topologies produced by the two different
methodologies but does not, as such, rule out the possibility of using one
of the nodes as a central node \cite{seli}.

In \cite{long}, we estimated the overall goodness of power-law fits by calculating the $R^2$ {\it coefficient of determination}, 
a measure which indicates the fraction of the total variation explained by the least-squares regression line. 
Averaged over all the time windows, we obtained the values $R^2 \approx 0.93$ and $R^2 \approx 0.86$, with and without outliers 
excluded, respectively. Since there were no outliers in the data for asset graphs, it was used as such to give an average 
of $R^2 \approx  0.75$. This indicates that scale free behavior is not evident in this case.

\section{Summary and conclusion}

In summary, we have introduced the concept of dynamic asset graph and
compared some of its properties to the dynamic asset tree, which we have
studied recently.  Comparisons between edge length distributions reveal
that the asset tree, due to the minimum spanning tree criterion, is forced
to ``accept'' edge lengths that are far less optimal (longer) than the
asset graph. This results in a higher average value for the normalized
tree length than for the normalized graph length although, in general,
they behave very similarly. However, the latter tends to exaggerate market
anomalies and, consequently, the normalized tree length seems to track the
market better. The asset graph was also found to exhibit clearly higher
single-step survival ratio than the asset tree. This is understandable, as
the spanning tree criterion does not allow the shortest connections, which
were conjectured to have the longest lives, to be included in the tree 
and, thus, their omission leads to a more fragile structure. This is also
witnessed by studying the multi-step survival ratio, where it was found
that in the early time horizon the asset graph shows exponential decay,
but the asset tree decays faster than exponential.  Later on, however, 
both decay as a power-law, but here the situation is reversed and the
asset tree decays more slowly than the asset graph. We also studied the
vertex degree distributions produced by the two alternative
approaches. Earlier we
have found asset trees to exhibit clear scale free behavior, but
for the asset graph scale free behavior is not so evident.  Further,
the values obtained for the scaling exponent are very different from our
earlier studies with asset trees.

\begin{acknowledgments}
J.-P. O. is grateful to European Science Foundation for REACTOR 
grant to visit Hungary, the Budapest University of Technology 
and Economics for the warm hospitality. Further, the role of Harri Toivonen 
at the Department of Accounting, Helsinki School of Economics, 
is acknowledged for carrying out CRSP database extractions. 
J.-P. O. is also grateful to the Graduate School in Computational 
Methods of Information Technology (ComMIT), Finland.
The authors are also grateful to R. N. Mantegna for very useful 
discussions and suggestions. This research was partially 
supported by the Academy of Finland, Research Center for 
Computational Science and Engineering, project no. 44897 
(Finnish Center of Excellence Program 2000-2005) and OTKA (T029985). 
\end{acknowledgments}


\begin{thebibliography}{10}
\bibitem{seli} The reason why the concept of central node 
is not applicable to dynamic asset graphs is that nothing guarantees that we have just one graph. We may have many 
that are not connected, and thus we cannot determine how far from each other these graphs are.
\end{thebibliography}
\end{document}